\documentclass[fleqn,usenatbib]{mnras}

\usepackage{newtxtext,newtxmath}

\usepackage[T1]{fontenc}

\DeclareRobustCommand{\VAN}[3]{#2}
\let\VANthebibliography\thebibliography
\def\thebibliography{\DeclareRobustCommand{\VAN}[3]{##3}\VANthebibliography}

\usepackage{graphicx}	
\usepackage{amsmath}	
\usepackage[dvipsnames]{xcolor}

\newcommand{\Lgal}{\textsc{L-Galaxies}}

\title[Correcting low-mass quiescent galaxies in SAMs]{Correcting for the overabundance of low-mass quiescent galaxies in semi-analytic models}

\author[J. E. Harrold et al.]{
Jimi E. Harrold$^{1}$, 
Omar Almaini$^{1}$,
Frazer R. Pearce$^{1}$,
Robert M. Yates$^{2}$
\\
$^{1}$School of Physics \& Astronomy, University of Nottingham, Nottingham, NG7 2RD, UK\\
$^{2}$Centre for Astrophysics Research, University of Hertfordshire, Hatfield, AL10 9AB, UK\\
}

\date{Accepted XXX. Received YYY; in original form ZZZ}

\pubyear{2023}

\begin{document}
\label{firstpage}
\pagerange{\pageref{firstpage}--\pageref{lastpage}}
\maketitle

\begin{abstract}
We compare the \Lgal{} semi-analytic model to deep observational data from the UKIDSS Ultra Deep Survey (UDS) across the redshift range $0.5 < z < 3$. We find that the over-abundance of low-mass, passive galaxies at high redshifts in the model can be attributed solely to the properties of `orphan' galaxies, i.e. satellite galaxies where the simulation has lost track of the host dark matter subhalo. We implement a simple model that boosts the star-formation rates in orphan galaxies by matching them to non-orphaned satellite galaxies at a similar evolutionary stage. This straightforward change largely addresses the discrepancy in the low-mass passive fraction across all redshifts. We find that the orphan problem is somewhat alleviated by higher resolution simulations, but the preservation of a larger gas reservoir in orphans is still required to produce a better fit to the observed space density of low-mass passive galaxies. Our findings are also robust to the precise definition of the passive galaxy population. In general, considering the vastly different prescriptions used for orphans in semi-analytic models, we recommend that they are analysed separately from the resolved satellite galaxy population, particularly with JWST observations reigniting interest in the low-mass regime in which they dominate.
\end{abstract}




\section{Introduction}
Our view of galaxy formation has evolved rapidly in recent decades, driven largely by deep observational surveys, which allow us to study large numbers of galaxies at different slices through cosmic history. To gain a deeper understanding of the underlying physics, it is also vital to compare our models of galaxy formation with these observations. A key test is to reproduce the galaxy stellar mass function (SMF) and its evolution.  When the SMFs between theory and observations agree over a wide range in redshift and stellar mass it provides a level of confidence that our understanding is likely to be broadly correct. In contrast, if we cannot explain the evolution of the SMF it may indicate a misunderstanding of the underlying physical processes. Further splitting SMFs by redshift, star formation rate (SFR), environment, galaxy class, etc., can help to diagnose likely problems or missing ingredients in our models \citep{henriques_galaxy_2015,croton_semi-analytic_2016,cattaneo_new_2017,cora_semi-analytic_2018, lagos_quenching_2023}. 
Large numbers of galaxies are required in order to directly compare the SMFs between observations and models accurately over a range of different parameters, and large  volumes of the Universe should be modelled to overcome the influence of cosmic variance \citep{somerville_cosmic_2004}. Until recently, the difficulty in obtaining a statistically significant sample of both simulated and observed galaxies across a range of environments and redshfits has proven to be prohibitive, preventing such direct comparisons.
\par

Wide-field deep near-infrared surveys provide a powerful method for probing high redshift ($z>1$) galaxy populations. Deep near-infrared studies are vital for probing rest-frame optical light at high redshift, and to ensure that we are not biased against old or dusty galaxies \citep{cimatti_old_2004}. In addition to infrared depth, a wide field is also required to accurately determine the space density of rare massive galaxies, and to study galaxy clustering and large-scale structure (e.g. \citealp{hartley_studying_2013}).
In particular, surveys such as the UKIDSS Ultra Deep Survey (UDS) and UltraVISTA can probe the SMFs of typical (sub-L$_\ast$) galaxy populations to redshifts of around $z=3$ (e.g., \citealp{muzzin_evolution_2013,taylor_role_2023}). Such surveys
are ideal for studying galaxy evolution around "cosmic noon" ($z=1-3$), at the peak epoch of both star formation and quenching (e.g. \citealp{ilbert_mass_2013}; \citealp{forster_schreiber_star-forming_2020}). Comparisons between observations and models of galaxy evolution at these epochs are therefore ideal for testing our understanding of galaxy formation and evolution.
\par

Semi-analytical models (SAMs) provide a potent method for comparing galaxy populations directly to observations \citep{white_galaxy_1991, springel_gadget_2001,baugh_primer_2006,somerville_physical_2015}. By opting to replace the full hydrodynamical prescriptions typically used to model a galaxy with simple analytical counterparts, and exploiting the massive computational speed gains achieved by doing so, SAMs can simulate substantial volumes and produce large catalogues of galaxies. These catalogues can be tuned to produce similar populations of galaxies to those observed in deep wide-field surveys such as the UDS, and thus can be directly compared against them on observables not included in the tuning process. Direct tracers of the underlying evolution of the galaxy population, such as the SMF, can then be used to test our contemporary understanding of galaxy evolution.
\par

While it has been shown that SAMs reconstruct a range of local galaxy properties with a reasonable degree of accuracy, historically they have found it harder to simultaneously tune to observations at higher redshifts \citep{asquith_cosmic_2018,fontanot_many_2009,weinmann_fundamental_2012,knebe_nifty_2015}. This difficulty is exacerbated as observational surveys probe to lower masses and even higher redshifts. Increasingly observations point to a so-called  `anti-hierarchical' form of growth, otherwise known as `mass assembly downsizing', where high mass galaxies formed rapidly in the early Universe, with low-mass galaxies forming at late times \citep{cimatti_mass_2006,lee_assembly_2013}.
Many SAMs, such as \Lgal{} have been tuned to reproduce the overall SMF across the redshift range $0.5 <z <3.0$. However, one particular tension between SAMs and observations still exists: SAMs overproduce the number of passive low-mass galaxies, with the overproduction becoming worse at higher redshifts \citep{asquith_cosmic_2018,donnari_quenched_2021}. Previous works such as \cite{donnari_quenched_2021} have shown that while some differences in the passive fraction can be attributed to the selection method used, this mainly alters the fraction at stellar masses above $\mathrm{10^{10.5}~M_\odot}$. Overall, it has become clear that SAMs overproduce the space density of low-mass passive galaxies by a factor of $\sim 3$ at low redshift, rising to up to over an order of magnitude more than is observed at high redshifts \citep{asquith_cosmic_2018}. In this letter we identify a significant subset of galaxies, orphan galaxies, that are causing this tension.
\par

SAMs are built on top of an underlying N-body simulation within which collapsed structures are identified using a halo finder and connected from one output to the next using a tree builder \citep{springel_populating_2001,behroozi_average_2013,srisawat_sussing_2013}.
In isolation the evolution and growth of dark matter haloes is relatively straightforward to follow. However, when haloes merge the density field can become difficult to disentangle. In particular, sub-haloes orbiting within a larger host dark matter halo can become lost and/or disrupted. Baryonic objects, where gas has cooled, collapsed and formed stars at the centre of these haloes can survive this process. Within SAMs, this process produces so-called orphan (type-II) galaxies \citep{hopkins_mergers_2010}. These are satellite galaxies where the underlying simulation has lost track of the host dark matter halo, either due its tidal disruption or simply falling below the resolution limit of the simulation. The treatment of orphan galaxies varies across different SAMs, and has been a topic of historical disagreement: some SAMs simply remove orphan galaxies completely, stating this improves the fit to the observational data used to tune the model \citep{croton_semi-analytic_2016}. A more common approach involves stripping away the gas supply to the galaxy and implementing a merger timescale between the central galaxy and the orphan \citep{springel_gadget_2001,guo_dwarf_2011}. This concept is physically motivated by the idea that without a dark matter halo the potential well of the galaxy is too shallow to retain gas against external forces such as ram-pressure stripping, which act on a short dynamical timescale \citep{mccarthy_ram_2008}. Older models such as \cite{de_lucia_hierarchical_2007} typically stripped all satellite galaxies (both orphans and resolved satellites) instantly. This idea has been revised in works such as \cite{guo_dwarf_2011}, where satellites strip slower and instantaneous stripping only applies to the hot gas that remains once a galaxy becomes an orphan. As this quenching happens rapidly these galaxies quickly become identified as quiescent. In the local universe this model appears to work well down to moderate masses, especially when compounded with the fact that it is difficult to separate orphans and satellites observationally. As observations have probed deeper and mass functions become complete to lower stellar masses the tension between the observed passive galaxy counts and the models has increased \citep{asquith_cosmic_2018,donnari_quenched_2021}. Some attempts to circumnavigate the orphan problem utilise the `excellent convergence' in numerous metrics between the two millennium simulations shown in works such as \cite{boylan-kolchin_resolving_2009}, employing a hybrid of two separate dark matter simulations. This solution was first attempted in \cite{guo_dwarf_2011}, which used Millennium-I for high and intermediate mass galaxies, combined with a run on Millennium-II for lower-mass galaxies. These results show improvements to the lower-mass passive fraction at the cost of increased runtime due to the use of merger trees from two cosmological simulations. Additionally, such models fail to directly address the underlying problem highlighted in this letter: the way orphans are currently modelled in SAMs is imprecise and there is a lack of consistency between the physics applied to orphans and resolved satellites. SAMs such as GAEA have attempted to address the low-mass passive population by slowing the quenching in satellites in works such as \cite{xie_influence_2020}, but still fail to rectify the issue (see \citealt{de_lucia_tracing_2024}; figure 11). \cite{luo_resolution-independent_2016} implemented a resolution independent treatment of orphans and satellites in an older version of L-Galaxies, finding an under-abundance of quenched central galaxies in the model (in addition to an over-abundance of quenched satellites), indicating that issues with quenching are not solely of environmental origin. However, adjustments made by \cite{luo_resolution-independent_2016} were performed on the 2013 version of \Lgal{}, which lacked the significant alterations to gas reincorporation and AGN feedback that the 2015 model provides, which aimed to address many of the discrepancies with central galaxies and how they quench \citep{henriques_galaxy_2015}. 
\par

In this letter, we apply \Lgal{} through a lightcone designed to mimic the properties of galaxies observed in the UKIDSS Ultra-Deep Survey (UDS), which is the deepest near-infrared ($K$-band selected) survey to date over such a large area of sky. By  separating galaxies into star-forming and passive subsets, we find that many of the discrepancies can be attributed  to the treatment of the orphan population in the models. We also show that by retroactively adjusting the properties of the orphan galaxies to follow a more gradual quenching prescription, the agreement between the model and survey dramatically improves, with a 4-fold reduction of the passive SMF $\chi^2$ values when compared with no adjustment. We show this result remains true when increasing the resolution of the underlying dark matter simulation, and when applying different standard cuts to define the passive fraction, including: specific SFR, UVJ cuts, and an \Lgal{} specific cut. We argue that while \Lgal{} is an excellent model for the luminosity evolution of the whole galaxy population it can be significantly improved for the passive population if the quenching of these galaxies is somewhat altered. We suggest that such a model should be implemented self-consistently into \Lgal{}. 
\par

Section 2 covers the usage of a catalogue derived from the 11th data release from the UKIDSS Ultra-Deep Survey, and how the standard output of \Lgal{} was processed in order to closely match the observational properties of this survey. Section 3 covers the direct comparison of galaxy SMFs as a result of this processing, highlighting the initial overabundance of low-mass passives, followed by implementation and comparison of the toy model. Section 4 discusses the conclusions we derive from applying this toy model, and possible routes for more permanent alterations to \Lgal{}. This letter assumes a flat $\Lambda$CDM cosmology of $\text{h}=0.7, \Omega_m=0.3, \Omega_\Lambda=0.7$; the standard cosmology of \Lgal{}  ($\text{h}=0.673, \Omega_m=0.315, \Omega_\Lambda=0.685$) has been scaled in the figures shown to match this cosmology.

\section{The Data}
\subsection{Observational Data}

For comparison to observations, we use a catalogue derived from the 
11th data release of the UKIDSS Ultra-Deep Survey (UDS; \citealt{lawrence_ukirt_2007}; Almaini et al., in preparation).  The UDS covers 0.77 square degrees,
reaching near-infrared depths of  $J=25.6$, $H=25.1$ and $K=25.3$. Additional imaging of the UDS in various wavebands has been performed using several instruments (all magnitudes given in AB, $5\sigma$ S/N over 2 arcsec diameter): Subaru Suprime-CAM  ($B=27.6$, $V=27.2$, $R=27.0$ and $z'=26.0$; \citealt{furusawa_subaruxmm-newton_2008}); CFHT MegaCam ($u'=26.75$); VISTA ($Y=24.5$; \citealt{jarvis_vista_2013}); Spitzer IRAC from SpUDS (PI: Dunlop) at $3.6\mu m $ and $4.5\mu m$, reaching magnitudes of $24.2$ and $24.0$ respectively, supplemented by 
deeper imaging from SEDS \citep{ashby_seds_2013}.
The full area covered by all 12 bands is 0.63 sq degrees, after the masking of artefacts and bright stars.

Photometric redshifts were calculated following the method of \cite{simpson_prevalence_2013}, using the  with EAzY code \citep{brammer_eazy_2008}, as described in \cite{taylor_role_2023}. The approach is based on the default EAzY setup, with 12 Flexible Stellar Population Synthesis (FSPS) SED components \citep{conroy_propagation_2010}, augmented by three simple stellar population (SSP) templates. These additional SSP templates, with ages of 20, 50, and 150 Myr, use a Chabrier IMF and sub-solar metallicity (0.2 solar) to represent recent bursts of star formation alongside the continuous star-formation histories in the FSPS templates. 
  Photometric redshifts were calibrated using 
approximately 8000 sources with secure spectroscopic redshifts, yielding a mean absolute dispersion of $\sigma_{NMAD}$ = 0.019 and an outlier fraction of  $\sim$ 3$\%$, where outliers are 
defined  when |$\Delta$z|/(1 + z) > 0.15. Further details of the photometric redshift determination can be obtained from \cite{taylor_role_2023}.

Stellar masses for both the UDS and the mock lightcone were calculated using the Bayesian analysis technique described in \cite{wild_evolution_2016}, based on a PCA analysis. A library of around 44000 stochastic burst model SEDs were built from \cite{bruzual_stellar_2003} stellar population synthesis models, accounting for a wide range metallicity, dust content, and star-formation histories.  When fitting to the model SEDs, this principal component analysis finds only three distinct eigenvectors are needed to reproduce over 99.7\% of the observed variance, with the 3 respective eigenvalues ("supercolours") being known as SC1, SC2, and SC3. Each of these SCs can be mapped to physical properties of the observed galaxies: SC1 correlates with the $R-$band weighted mean stellar age of a galaxy and with dust content; SC2 correlates with the fraction of stars formed in the last Gyr, and helps to break the degeneracy between dust and age;   SC3 helps to break the degeneracy between the fraction of stellar mass formed and metallicity seen in SC2. Galaxies are classified using their position in SC space to separate star-forming, dusty, and passive galaxies, in addition to rarer classes such as post-starburst galaxies. This technique has been verified spectroscopically
\citep[e.g. see][]{maltby_identification_2016,wilkinson_starburst_2021}, finding very low rates of contamination of star-forming galaxies in the passive class. The resulting separation of passive and star-forming galaxies is also in very good agreement with the UVJ rest-frame technique \citep{wuyts_what_2007}, as demonstrated in 
\cite{wild_new_2014} and \cite{almaini_massive_2017}. 
The influence of the precise selection method will be discussed further in section \ref{alterations}. We estimate stellar mass errors are typically $\pm 0.1$ dex  \citep[see][]{almaini_massive_2017}.  The same PCA technique is used to classify galaxies and determine stellar masses for the simulated galaxies within the lightcone (see below).

Stellar mass completeness limits at a given redshift are calculated using the method described in \cite{pozzetti_zcosmos_2010} using a 95\% completeness limit. We impose an additional minimum stellar mass limit of $\mathrm{\log(M_\ast/M_\odot)=9.5}$ to both the UDS and the mock galaxies, reflecting the minimum mass resolution of \Lgal{} when run on the Millennium simulation \citep{henriques_galaxy_2015}.

\subsection{Simulated Data}
The model used to compare to observations was obtained using the 2015 version of the Munich Galaxy Formation Model known as \Lgal{}, described in \cite{henriques_galaxy_2015}. The underlying DM simulation used is the Millennium simulation \citep{springel_simulations_2005}, with the cosmology retroactivly re-scaled to the standard UDS cosmology of $\text{h}=0.7, \Omega_m=0.3, \Omega_\Lambda=0.7$.  While this re-scaling does not fully account for adjustments detailed in \cite{angulo_one_2010}, we argue any changes to our conclusions as a result of this will be negligible. We use standard input parameters described in \cite{henriques_galaxy_2015} and originally constrained via a Markov chain Monte Carlo approach. Additional dust attenuation was included via the prescription described by \cite{charlot_simple_2000}. On top of the \Lgal{} output we construct a lightcone. This lightcone is designed to cover a field-of-view of 1 square degree over a redshift range of $0.5 < z < 3.0$. This closely matches the field-of-view of the UDS survey. Galaxies along the lightcone were then selected by sampling their data from the simulation snapshot that best matched the redshift at which they are being sampled. To prevent galaxies from being detected multiple times, the progenitor and descendants of a galaxy are marked in order to ensure they are not  re-used in the construction of the lightcone. A scatter of 0.08(1 + z) was applied to the stellar mass values before processing, derived from \cite{conroy_propagation_2009}, to represent the systematic uncertainty in observational stellar mass determination. An additional scatter of 0.0189(1 + z) is applied to the redshift values of the catalogue, reflecting the estimated uncertainty derived by comparing spectroscopic and photometric redshift values for sources within the UDS \citep{wilkinson_starburst_2021}.

Galaxy SEDs were generated in a similar way to our observational sample, by matching each galaxy to a library of around 44000 stochastic burst model SEDs built from \cite{bruzual_stellar_2003} stellar population synthesis models, accounting for a wide range of metallicities, dust content, and star-formation histories. Each galaxy SED is then convolved with the 13 UDS filters at various redshifts, stepping by $\Delta z=0.1$ through a redshift range of $0.5 < z < 3.0$, giving expected flux values for the mock galaxy as if it was observed at each of the various redshifts. Using the 13 flux values generated for each mock galaxy, the lightcone is then run though the same principal component analysis described above.  
\par

Once the SC values have been generated, stellar masses and SFRs can be calculated for the galaxy using the technique outlined in \cite{wild_evolution_2016}. All properties then calculated are obtained in precisely the same way as the real catalogue. We find our constructed lightcone to be in strong agreement with typical \Lgal{} output and production tests.

\begin{figure*}
    \includegraphics[width=0.99\textwidth]{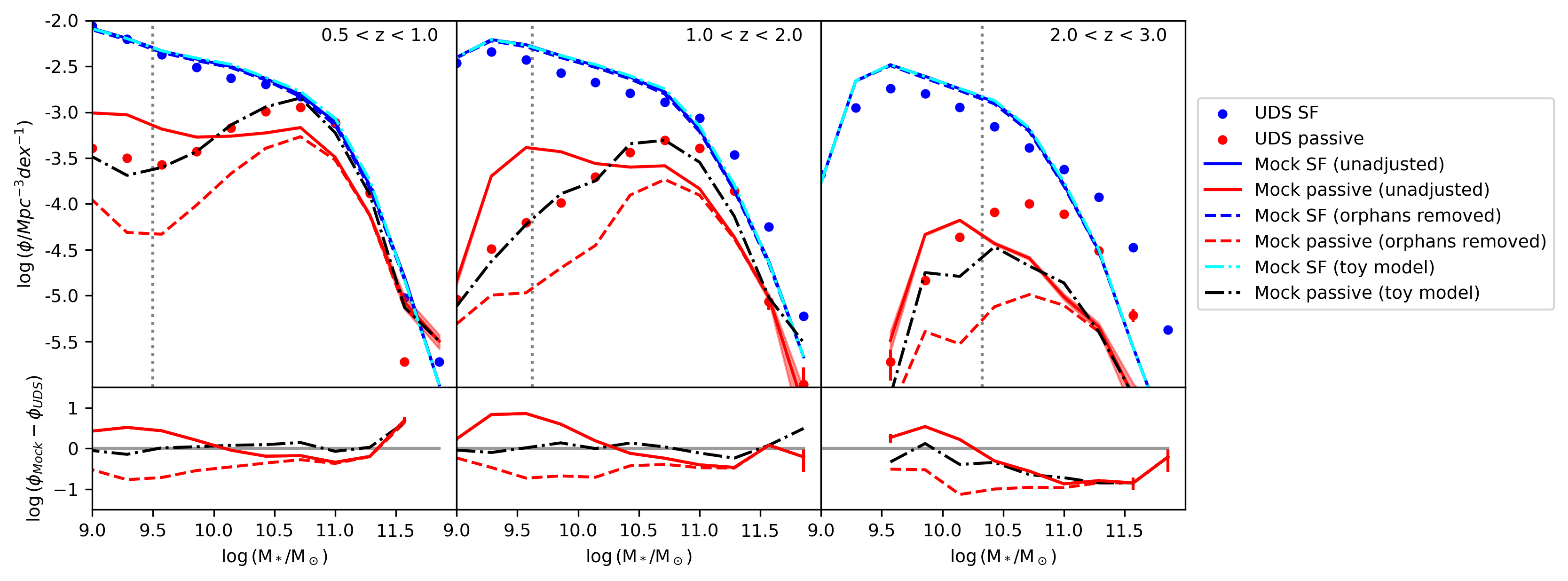}
    \caption{The mass functions of star-forming (blue) and passive (red) galaxies in three redshift ranges. The observational data from the UDS is shown by the solid dots. Three variants of the \Lgal{} model output are shown:  unadjusted (solid lines), orphans removed (dashed lines), and our toy model with modified orphans (black, dot-dashed lines). The vertical line denotes the 95\% completeness limit, determined using the method of \protect\cite{pozzetti_zcosmos_2010}. The bottom row shows the difference between the models and the UDS mass functions.}
    \label{fig:mock_environment}
\end{figure*}

\section{Comparison of galaxy mass functions}
\begin{figure}
\center
\includegraphics[width=0.95\columnwidth]{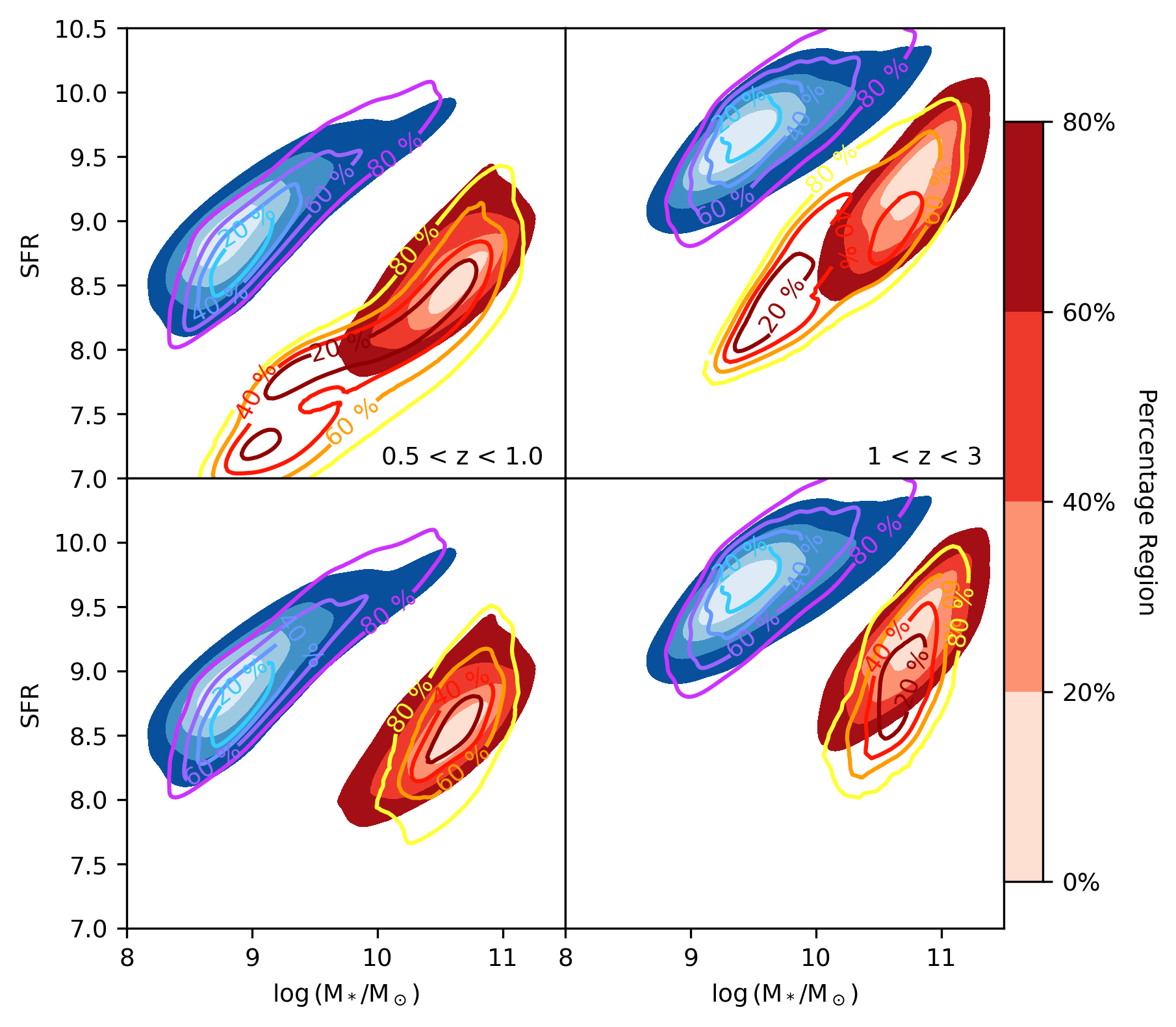}
   \caption{Contours for star formation rate (SFR) for both the star forming (blue) and passive (red) populations. In all four panels the observational data is displayed as the background shading. The contours are taken from the models, respectively \Lgal{} unaltered (first row), and \Lgal{} with the modified orphans (second row). Similar to the SMFs, the star forming contours show very little change when adjusting for orphans, however the passive contours show a significant improvement in fit.  }
   \label{fig:fixed_gal}
\end{figure}

We compare the galaxy SMFs produced by the mock lightcone and the UDS across the redshift range $0.5<z<3.0$ in Figure~\ref{fig:mock_environment}. Galaxies are split into star forming and quiescent bins via a supercolour cut defined by \cite{wild_new_2014} and further refined in \cite{wilkinson_starburst_2021}. The SMF for the mock is generated both with and without orphans included. Excluding the orphans from the star-forming SMF (blue lines) makes no appreciable difference across all redshifts, while the quiescent SMF (red lines) loses the large tail of low-mass galaxies seen in both upper panels. While removing the orphans shows an improvement in the overall shape of the quiescent SMF, the density of low-mass passive galaxies drops by as much as an order of magnitude below that seen in the UDS. This drop is mostly due to the fact that orphans make up at least 35\% of the total quiescent population produced by \Lgal{}. We explore the effects of various changes in selection method, the DM model used, and the version of \Lgal{} used in section \ref{alterations}
\par

While removing orphans improves the general shape of the SMF, the loss of around a third of the quiescent population itself produces problems: 
the low-mass passive population is now under- rather than over-represented. While this underproduction could be addressed by re-calibrating \Lgal{} , we argue that simply removing orphan galaxies without a physical mechanism is likely to be problematic. 
\par
\subsection{Feeding orphans}
\label{feed}
In an attempt to address this point we implement a toy model that aims to rectify the issues presented above without requiring new modules and re-calibration of \Lgal{}. Our model aims to substitute orphan galaxies with higher SFR and higher mass satellite galaxies which have otherwise comparable properties. This is motivated by the idea that while the growth of the disc region of a galaxy closely follows the growth of the host halo (and thus is interrupted when this halo is no longer present in an orphan) the central regions of the galaxy present a deep enough potential well to retain some of the gas reservoirs that are currently considered stripped. This allows star formation to continue for longer as the would-be orphan falls into the central cluster. This model has some observational motivation, e.g. \cite{cortese_dawes_2021} discuss that stripping is not sufficient to completely remove cold gas discs in cluster galaxies above masses of $\mathrm{M_*>10^8 M_\odot}$, allowing star formation to continue centrally for several Gyr. \cite{bamford_sizes_2007} found star formation is more central in satellite galaxies compared to their field counterparts, with \cite{weinmann_environmental_2009} finding little evidence for rapid quenching effecting satellites, preferring a gradual quenching path created by starvation. \cite{kawinwanichakij_effect_2017} finds a "lack of significant environmental quenching of low mass galaxies at $z>1$", supporting a model where gas is depleted through over-consumption and outflows as opposed to environmental effects, but note that effects such as disk fading are needed to explain the transition from star forming to quiescent. \cite{woo_satellite_2017} also notes a consistent increase in the surface density of satellites as they transition through the green valley, suggesting either an enhancement of star formation in the central regions or quenching effects in the outermost regions of a galaxy. Allowing the continuation of star formation in the models will have three key effects on the overall galaxy population: (1) many of the low-mass orphans will be moved to the star forming region of the SMF, alleviating the overproduction of passive galaxies while having little impact on the star-forming SMF (since star-forming galaxies are far more numerous); (2) once the orphans do quench, they do so at a higher mass, which acts to help build the 'knee' of both the star forming and quiescent SMFs; (3) low-mass galaxies with low or decreasing halo mass should retain more star formation and grow past the typical halo mass to stellar mass relationship, allowing slightly higher-mass galaxies to exist before stripping and reclassification as orphans takes place.
\par

In order to provide adjusted properties for these modified orphans we use the properties of satellite galaxies with equivalent redshifts and supercolour values. Galaxies were binned using a k-d tree to ensure a minimum population of equivalent satellite galaxies in each colour bin without sacrificing binning resolution in high density regions. Using standard 2d-histogram bins over a range of bin widths produces similar results. Using these bins, the \Lgal{} output is modified; each orphan has a random satellite substitute selected from within the same bin. The properties of this satellite are then used in place of the orphan's when processing the lightcone. This simplistic approach allows orphans to assume the properties of galaxies undergoing similar expected evolutionary pathways. Using different properties to initially bin the galaxies such as $U-V$ and $V-J$ rest-frame colours, has little effect on the conclusion. Further robustness tests are presented in section \ref{alterations}.

The result of feeding the orphans can be seen in the lower panels of Figures~\ref{fig:mock_environment} \& \ref{fig:fixed_gal}. We observe an improvement in the overall fit of the passive fraction across all redshifts, while having a negligible impact of the star forming SMFs. SFRs show a similar improvement, matching closely with the SFRs produced by the UDS observations. 
\par

This improvement of the quiescent fraction may complement recent works such as those of \cite{murphy_l-galaxies_2022}, who employ gradual tidal stripping of the cold gas supplies of satellites, including orphans, improving the overall fit to the observations used. This adjustment differs from our proposal, which primarily effects the size of available gas reservoirs before tidal stripping takes place, but similarly promotes a less violent treatment of orphan galaxies. While no direct comparison of the passive fraction is made in the paper, it is likely that the low-mass passive fraction will increase under the \cite{murphy_l-galaxies_2022} prescription, due to an earlier onset of cold gas stripping in satellites, exacerbating the tension. Combining such a model with our prescription, where orphans have access to a larger quantity of gas to allow for boosted levels of star formation, may allow the stated improvements while still reproducing the passive fraction at high redshifts.
\par

\section{Robustness Tests}
\label{alterations}

\subsection{Quiescent Selection}
In order to ensure that our primary conclusions are a robust feature of the model and not an artifact of selection methods, we apply several different quiescent selection techniques to the mock lightcone and the observational data. We find a similar overproduction due to orphans when selecting via a standard sSFR (e.g. $\log(\text{sSFR}/\text{yr}^{-1})<-11$), UVJ colour selection (e.g., as outlined in \citealt{wuyts_what_2007}, \citealt{muzzin_evolution_2013}), and the adjusted UVJ technique outlined in \cite{henriques_galaxy_2015}. We note that the discrepancies we find largely agree with those found in section 5.1 of \cite{henriques_galaxy_2015}, which also notes an excess of low-mass passive galaxies at increasing redshifts despite calibrating to the passive fraction. This problem may be exacerbated in our results due the increased selection accuracy of the supercolour technique, combined with the close matching of the mock data to our observational data set.

We also compared our results with observational data from the
COSMOS/UltraVISTA deep survey \citep{wilkinson_starburst_2021} at $z<1$, finding consistent results and the same discrepancy in the predicted fraction of low-mass passive galaxies.

\subsection{Model Selection}
 We also obtain similar results if SMFs are generated without using a lightcone, and are instead produced directly from individual snapshots with observational limits applied. The 2020 version of \Lgal{} \citep{henriques_l-galaxies_2020} also shows the same problem and subsequent improvement, which is to be expected as the underlying treatment of orphans did not change.
\subsection{Millenium-II}
Our findings are also independent of the details of lightcone construction and the particular version of \Lgal{} used. As orphans are a resolution dependent phenomenon, it is expected that the effect would be lessened in higher resolution simulations. We confirmed this by running partial analysis on halo trees produced by Millennium-II \cite{boylan-kolchin_resolving_2009}, which provides 5 times better spatial resolution, down to a softening length of 1 kpc/h, and 125 times better mass resolution of $\mathrm{6.9\times10^6\, h^{-1}M_\odot}$. We find the results produced by Millennium-II show an improved initial shape of the passive galaxy SMF, but the low-mass tail was still improved with orphan galaxies removed, suggesting that while orphans are indeed a resolution dependent feature, their problematic prescriptions persist. Due to the precise integration of the lightcone with \Lgal{} 2015 and the requirement for large simulation box sizes to accurately capture the complete cosmology represented in the observations used, it was not possible to run a direct lightcone comparison to Millennium-II. Instead we apply a $K$-band magnitude cut at $25.3$, matching the $5\sigma$ $K$-band limit from data release 11 of the UDS, to act as a basic proxy to the more sophisticated detection criteria imposed by the lightcone.
\par
\subsection{Selection of satellite replacements}
In our simple model we match orphans to alternative non-orphaned 
satellite galaxies with similar redshifts and PCA parameters. We investigated the impact of additional matching to (i) the distance of the satellite galaxy from the central galaxy, (ii) the virial halo mass before infall, and (iii) the SFH before infall. We find no significant changes to the resulting mass functions. This is likely due to our focus on quiescent galaxies which, by definition, show little change in mass once quenched and thus do not vary significantly in their properties with parameters such as distance from the central. Additionally, many of these parameters are partially fit simply by fitting to the PCA parameters before infall, which will likely account for secondary dependencies on properties such as halo mass.
\par

\section{Conclusions}
We construct a mock lightcone from the semi-analytic model \Lgal{}, designed to closely match the observational constraints of a galaxy population derived from the UKIDSS UDS. We show that this mock lightcone overproduces low-mass passive galaxies and that this excess is robust to changes in the mock catalogue construction process.
\par

In the context of a semi-analytic model, orphan galaxies are defined as those galaxies where the surrounding dark-matter halo can no longer be tracked. Typically the host halo has been stripped as the galaxy orbits within a larger galaxy cluster or group. In these cases the galaxy itself may survive but its location is no longer known accurately. By isolating the orphan galaxy population we show that the majority of the discrepancy seen in the stellar mass functions across our redshift range of $0.5<z<3.0$ can be explained by an improper treatment of this population. We show that this result holds regardless of the choice of the underlying simulation, the selection method used to identify quiescent galaxies, or the version of \Lgal{} used.
\par
We argue that a more realistic treatment of orphan galaxies can be obtained within semi-analytic models, such as \Lgal{}, by not quenching their star formation as rapidly. Our suggestion is that the central regions of these galaxies should retain an increased gas reservoir, either directly via a larger supply of cold gas, or indirectly by retaining a hot gas reservoir out to the virial radius for longer than the current stripping prescription allows. By producing a toy model implementing this idea, we demonstrate a version of \Lgal{} that shows a remarkable alignment with the observed passive galaxy SMFs across the entire studied redshift range out to $z=3$.
\par
Within the complex and challenging environment of a galaxy cluster the extensive dark matter haloes of galaxies are naturally stripped away and merged into the general halo of the cluster itself. In such an atmosphere some galaxies will lose their entire surrounding dark matter envelope. Within a simulation a galaxy's host dark matter halo can disappear and be untraceable. What to do with the galaxy in these cases is an ongoing and well-known problem for semi-analytic modellers and there have been a variety of solutions implemented. This population exists in all semi-analytic models, even if it is sometimes neglected. The low-mass end of the passive population is particularly sensitive to this orphan population and alignment with observational results is much improved if some ongoing residual star formation is allowed.

\par

\section*{Acknowledgements}
This work was supported by the Science and Technology Facilities Council grant numbers ST/X000982/1 and ST/X006581/1. We thank David Maltby, Will Hartley, Kate Rowlands, and Vivienne Wild for valuable discussions.

\section*{Data Availability}
The data specifically shown in this paper will be shared upon request to the corresponding author. 



\bibliographystyle{mnras}
\bibliography{references} 





\bsp	
\label{lastpage}
\end{document}